\documentclass[doublecol]{epl2} 

\usepackage{amsmath} 
\usepackage{amssymb} 
\usepackage{cite}
\usepackage{subfigure} 
\usepackage{hyperref} 	
\usepackage{color}

\newcommand{\kk}{\mathbf{k}} 

\newcommand{\rr}{\mathbf{r}}

\newcommand{\vv}{\mathbf{v}}

\newcommand{\EE}{\mathbf{E}}
\newcommand{\AAA}{\mathbf{A}}

\newcommand{\eq}[1]{(\ref{#1})}

\title{Berry curvature effects in the Bloch oscillations of a quantum particle under a strong (synthetic) magnetic field}

\shorttitle{Bloch oscillations under a strong magnetic field} 

\author{Marco Cominotti \inst{1,2} \and Iacopo Carusotto \inst{3}}
\shortauthor{Cominotti {\em et al.}}

\institute{                    
  \inst{1} Universit\`a di Trento,  Dipartimento di Fisica, via Sommarive 14, 38123 Povo, Italy\\
  \inst{2} Universit\'e Grenoble 1/CNRS, Laboratoire de Physique et de Mod\'elisation
des Milieux Condens\'es (UMR 5493), B.P. 166, 38042 Grenoble, France\\
  \inst{3} INO-CNR BEC Center and Dipartimento di Fisica, Universit\`a di Trento, via Sommarive 14, 38123 Povo, Italy\\
}


\abstract{ We study the magnetic Bloch oscillations performed by a quantum particle moving in a two-dimensional lattice in the presence of a strong (synthetic) magnetic field and a uniform force. An elementary derivation of the Berry curvature effect on the semiclassical trajectory is given as well as an explicit connection to the classical Hall effect in the continuum limit. Perspectives to observe these effects in optical systems using synthetic gauge fields for photons are discussed.}

\begin{document}

\maketitle

In spite of its simple formulation, the coherent dynamics of a single quantum particle trapped in a periodic lattice potential and subject to a strong magnetic field displays a remarkably rich physics and is responsible for several important effects of condensed-matter physics.
While the weak magnetic field case is successfully described by the semiclassical theory of Bloch electrons~\cite{ashcroft}, magnetic field strengths on the order of one quantum of magnetic flux per unit cell of the lattice are responsible for qualitatively different behaviours: the Bloch bands of the lattice are split into many subbands with a fractal structure in the energy vs. magnetic field plane, the so-called {\em Hofstadter butterfly}~\cite{hofstadter}, and the non-trivial topology of the Hofstadter bands requires that the Berry curvature is included in the semiclassical theory of transport~\cite{niu,sundaram}.

While many consequences of this physics are nowadays known and well understood at a macroscopic level, much less experimental literature is available on microscopic studies of the basic mechanisms: in ordinary solids, huge magnetic fields are required to observe this physics and many spurious decoherence phenomena may compete with the quantum transport. Only recently, the experimental advances in the generation of strong {\em synthetic gauge fields} for (neutral) atoms~\cite{dalibard} and for photons~\cite{rmp} are opening new directions in the study of the quantum dynamics under strong magnetic fields: first examples in this direction are the nucleation of quantized vortices under the effect of a synthetic gauge field in an atomic condensate~\cite{linb} and the peculiar light propagation effects in the presence of a synthetic gauge field for photons that have been observed in deformed honeycomb-shaped waveguide arrays~\cite{rechtsman,rechtsman2} and in suitably designed arrays of coupled optical cavities~\cite{hafezi}.

In the present Letter we provide an unified and elementary derivation of the semiclassical theory of transport including the effect of the Berry curvature of the Bloch bands~\cite{niu,sundaram}, and we apply it to study the Bloch oscillations performed by a quantum particle trapped in a two-dimensional lattice and subject to a strong synthetic magnetic field when a uniform and time-independent force is applied to it. 
While scattering and dephasing effects make the observation of Bloch oscillations (BO) of electrons in ordinary solids extremely difficult~\cite{superlattices}, a neat observations of atomic BO in optical lattices has been demonstrated~\cite{salomon,roati,cristiani} and very recently exploited to map the structure of atomic Bloch bands in a honeycomb-like lattice potential~\cite{graphene_atoms}; theoretical studies of the effect of the Berry curvature, of diverse origin, onto atomic BO in different geometries have appeared in~\cite{dudarev,cooper,modugno_berry}.

Beside atomic systems, a most promising alternative to study this physics is nowadays offered by optical systems, where BO of photons have been experimentally observed in waveguide arrays \cite{pertsch,morandotti} and in periodic dielectric systems~\cite{sapienza}. The present Letter will hopefully stimulate experiments in the direction of investigating the effect on the BO of the strong synthetic gauge fields for photons demonstrated in~\cite{rechtsman,hafezi,rechtsman2}.


\section{The model}
We consider a particle confined in a two-dimensional square lattice of spacing $a$, immersed in a uniform synthetic magnetic field $\textbf{B}=B\hat{z}$ perpendicular to the $xy$-plane of the lattice~\footnote{As we are dealing with synthetic gauge fields, only the products $e\EE$ and $e\bf{B}$ have a physical meaning.}. The vector potential associated to the uniform magnetic field $\textbf{B}$ in the Landau gauge has the form $\textbf{A}(\textbf{r})=Bx\hat{y}$.

Within a standard tight-binding (TB) approximation for the periodic lattice potential, the single particle Hamiltonian can be written in the Peierls form 
\begin{equation}
H=-J\sum_{\langle i,j\rangle}\hat{a}^{\dagger}_{i}\hat{a}_{j}e^{i\phi_{ij}}\,,
\label{eq:tbh}
\end{equation}
as the sum of hopping terms. The indices $i$ and $j$ indicate the lattice site, and the sum is restricted only to the nearest-neighboring sites. The $\hat{a}^{\dagger}_{i}$ ($\hat{a}_{i}$) operators create (annihilate) one particle at the lattice site $i$ and satisfy usual bosonic commutation rules $[\hat{a}_{i},\hat{a}^{\dagger}_j]=\delta_{i,j}$. 
The positive coefficient $J$ quantifies the strength of tunneling from site $j$ to site $i$, while the hopping phase $\phi_{ij}$ can be written in terms of a line integral of the vector potential along the hopping path $\textbf{r}_j\rightarrow \textbf{r}_i$,
\begin{equation}
\phi_{ij}=-\frac{e}{\hbar}\int_{\textbf{r}_j}^{\textbf{r}_i}\textbf{A}(\textbf{r})\cdot \upd\textbf{r}\,.
\label{eq:phase}
\end{equation}
While the hopping phase $\phi_{ij}$ along a single link of the lattice is manifestly not gauge invariant, the sum of the phases accumulated while hopping around a closed contour is gauge invariant and proportional to the magnetic field flux encircled by the contour in units of the flux quantum $\phi_0=e/2\pi\hbar$. The key parameter determining the properties of the model is then $\alpha$, defined in terms of the number of magnetic flux quanta across the elementary cell of the lattice as
\begin{equation}
\alpha= \frac{\phi}{\phi_0}={-}\frac{1}{2\pi} \sum_{\Box}\phi_{ij}=\frac{e}{2\pi\hbar}Ba^2\,.
\end{equation}
In the Landau gauge used in this work, the hopping amplitude along the $\hat{x}$-direction has a phase $\phi_{ij}^{\hat{x}}=0$, while the one along the $\hat{y}$-direction has a phase $\phi_{ij}^{\hat{y}}=-eB ax/\hbar$.

\section{The band structure}
Even though the application of a uniform magnetic field does not affect the (discrete) translational symmetry of the physical system, this is not apparent in the Hamiltonian \eq{eq:tbh} where the hopping amplitudes have a spatially dependent phase stemming from the spatial dependence of the vector potential $\textbf{A}$. Of course, this breaking of the translational symmetry is only fictitious, as it is exactly compensated by a suitable gauge transformation.

To diagonalize the Hamiltonian \eq{eq:tbh}, it is useful to consider the so-called {\em magnetic translation group}~\cite{landau,zak}, whose operators $\tilde{T}_{\textbf{a}}$ are defined as 
$\tilde{T}_{\textbf{a}}=e^{-i e\Delta \textbf{A}_{\textbf{a}}(\textbf{r})\cdot\textbf{r}/\hbar }\,T_{\textbf{a}}$, in terms of the usual lattice translation operators $T_{\textbf{a}}$ and of the vector potential variation $\Delta\textbf{A}_{\textbf{a}}(\textbf{r})=\textbf{A}(\textbf{r}+\textbf{a})-\textbf{A}(\textbf{r})$. All the $\tilde{T}_{\textbf{a}}$ commute with the Hamiltonian $H$ in \eq{eq:tbh} but they do not in general commute among them. For instance, the magnetic translation operators for the two unit lattice vectors $\textbf{a}_{x,y}$ in the $\hat{x}, \hat{y}$-directions satisfy $\tilde{T}_{\textbf{a}_x}\tilde{T}_{\textbf{a}_y}=\tilde{T}_{\textbf{a}_y}\tilde{T}_{\textbf{a}_x}e^{-2\pi i \alpha}$, commutation of $\tilde{T}_{\textbf{a}_x}$ and $\tilde{T}_{\textbf{a}_y}$ is satisfied only if the magnetic flux through a plaquette is an integer multiple of the flux quantum $\phi_0$. On the other hand, for a rational $\alpha=p/q$ (with $p,q$ co-prime integers), it is possible to define a larger {\em magnetic unit cell} as the union of $q$ adjacent lattice plaquettes: the magnetic flux through the magnetic unit cell is then an integer multiple of $\phi_0$, but the wider spatial periodicity corresponds to a $q$ times smaller magnetic Brillouin zone (MBZ).
While the choice of the magnetic unit cell is not unique in general, we can take for simplicity a magnetic unit cell formed by $q$ adjacent plaquettes along the $\hat{x}$-direction. With this choice, the MBZ becomes of rectangular shape $[-\pi/qa,\pi/qa]\times[-\pi/a,\pi/a]$, with a shorter size along $\hat{k}_x$. 

Since the Hamiltonian $H$ in \eq{eq:tbh} commutes with the magnetic translation operators, we can look for the common eigenstates, which we will call {\em magnetic Bloch states}. This was first done by Hofstaedter~\cite{hofstadter} who reduced the problem to the solution of a one-dimensional difference equation, the so-called Harper equation:
\begin{equation}
-J\left[g_{m+1}+g_{m-1} - 2\cos(2\pi \alpha m-k_{y})g_{m}\right] =\mathcal{E}g_{m}\,.
\label{eq:harperm}
\end{equation}

Here, the spatial coordinates $x$ and $y$ have been discretized on the lattice points as $x=ma$, $y=na$, with $m,\,n\in\mathbb{Z}$. Thanks to the discrete translational symmetry along the $\hat{y}$-direction of the vector potential in the chosen Landau gauge, a plane wave form can be assumed for the eigenfunction $\psi_{m,n}$ along the $\hat{y}$-direction, $\psi_{m,n}=\exp(ik_{y}n)g_m$.

Even though the eigenvalues of the Hamiltonian \eq{eq:tbh} arrange in a very complex fractal structure in the energy $E$ vs. magnetic field parameter $\alpha$, the so-called {\em Hofstadter butterfly}, exciting physics is obtained in the case of a rational $\alpha=p/q$ where the effective potential appearing in the Harper equation \eq{eq:harperm} is periodic in the $\hat{x}$-direction with a period of $q$ lattice points, in agreement with the magnetic translation arguments. 

\begin{figure}[h]
\begin{center}
\subfigure{
\includegraphics[width=0.155\textwidth]{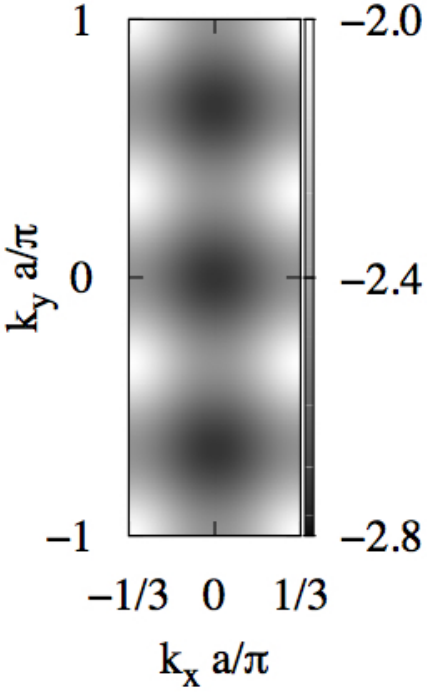}}
\subfigure{
\includegraphics[width=0.155\textwidth]{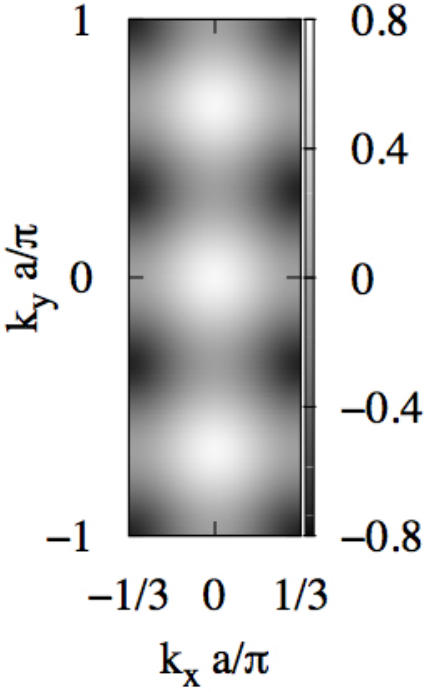}}
\subfigure{
\includegraphics[width=0.147\textwidth]{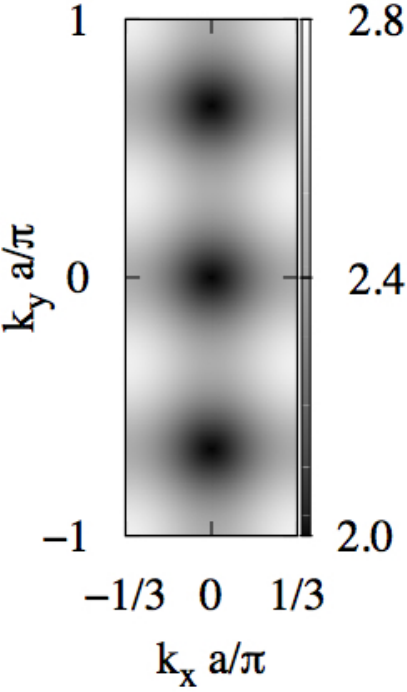}}
\caption{Energy dispersion of the three magnetic bands in the $\alpha=1/3$ case in units of the tunneling energy $J$.}
\label{fig:band13}
\end{center}
\end{figure}

In particular, for each value of $k_y$, the eigenstates of the Harper equation are then classified according to the Bloch theorem in terms of a $k_x$ wavevector contained in $[-\pi/qa,\pi/qa]$. As a result, we can classify the eigenstates of the Hamiltonian \eq{eq:tbh} in terms of $k_{x,y}$ within the magnetic Brillouin zone and plot the resulting magnetic bands. In the non-magnetic $\alpha=0$ case, one recovers the standard cosine shape of the tight-binding energy band, $\mathcal{E}=-2J\left[\cos(k_{x})+\cos(k_{y})\right]$. In the $\alpha=1/3$ case, the three bands have a non-trivial structure as shown in Fig.~\ref{fig:band13}.

\noindent  In particular, note the additional $q$-fold periodicity of the Hofstadter bands in the $y$ direction.

For decreasing values of $\alpha=1/q$, the standard physics of the Landau levels is recovered in that the $q$ bands become flatter and flatter and the spacing of the lower ones tends to the cyclotron frequency $\omega_c=eB/m$. This behavior can be recovered from the Harper equation \eq{eq:harperm} as follows: for $q\to \infty$, the minima of the oscillating potential becomes widely separated in space, so tunneling between them is suppressed. As the oscillating potential becomes smoother, the spacing between the eigenstates decreases and the lowest states end up exploring the harmonic bottom only. More details on this physics will be discussed in the last section of the Letter in connection with the classical Hall effect.
 
\section{Semiclassical dynamics}

In the previous section we have studied the band structure of the eigenstates of the particle motion in the presence of a uniform magnetic field. We now proceed to study the actual trajectory of a particle in space. Within a semiclassical picture, this will be done in terms of the time evolution of a wave packet contained within the $n$-th magnetic band and localized (as much as the Heisenberg principle allows it) in both real and $\textbf{k}$ space.

According to the modern semiclassical theory of electron transport in solids~\cite{niu,sundaram}, the time evolution of the wavepacket center of mass $\textbf{r}_c$ and $\textbf{k}_c$ is ruled by the following pair of equations,
\begin{subequations}
\begin{align}
\hbar\dot{\textbf{k}}_{c}(t)&=e\textbf{E}\,,\\
\hbar \dot{\textbf{r}}_{c}(t)&=\nabla_{\kk} \mathcal{E}_{n,\textbf{k}}-e\textbf{E}\times\mbox{\boldmath$\Omega$}_{n}(\textbf{k})\,.
\end{align}
\label{eq:semiclassicalberry}
\end{subequations}
The first equation, (\ref{eq:semiclassicalberry}a), describes the usual evolution of the (quasi-)momentum under the effect of the electric force $e\textbf{E}$, as discussed in solid-state textbooks~\cite{ashcroft}. Of course, accuracy of the semiclassical theory requires that this force is weak enough not to induce Landau-Zener interband transitions.
The second equations, (\ref{eq:semiclassicalberry}b), is much more interesting: while the first term recovers the usual group velocity $\vv^{\rm gr}_{n,\textbf{k}}=\frac{1}{\hbar}\nabla_{\textbf{k}}\mathcal{E}_{n,\textbf{k}}$ of a particle living on the energy band $\mathcal{E}_{n,\textbf{k}}$, the second term involves the so-called Berry curvature of the band, geometrically introduced as the curvature of the Berry connection,
\begin{equation}
\mbox{\boldmath$\Omega$}_{n}(\textbf{k})=\nabla_\kk\times \mathcal{A}_{n,\kk}= \nabla_\kk\times [i \langle u_{n,\textbf{k}}|\nabla_{\textbf{k}}u_{n,\textbf{k}}\rangle]\,,
\label{eq:curvberry}
\end{equation}
defined in terms of the periodic part $u_n(\textbf{k})$ of the Bloch wavefunction
$\psi_{n,\textbf{k}}(\textbf{r})=\exp(i\textbf{k}\cdot\textbf{r})\,u_{n,\textbf{k}}(\textbf{r}).$
This second contribution is the so-called anomalous or Berry velocity, which provides a spatial drift proportional to the temporal variation of the $\textbf{k}$ vector, and directed orthogonally to the applied force. In the solid-state context, this extra contribution to the velocity of the electrons was first noticed by Karplus and Luttinger \cite{karplus} in the framework of a linear response analysis. More recently, it has been reinterpreted in a more general framework as a Berry phase effect in \cite{niu,sundaram}. An elementary derivation is sketched in the next section.

\begin{figure}[t]
\begin{center}
\subfigure{
\includegraphics[width=0.144\textwidth]{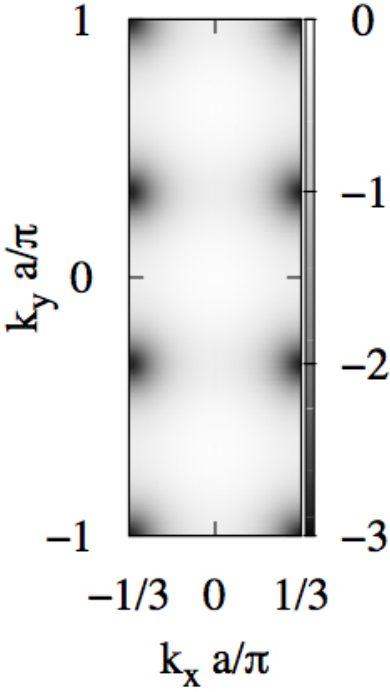}}
\subfigure{
\includegraphics[width=0.140\textwidth]{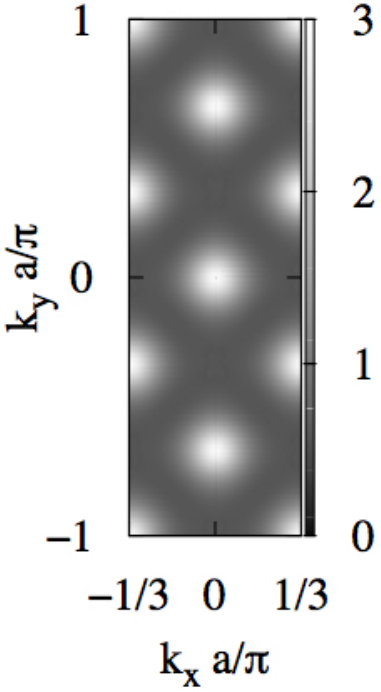}}
\subfigure{
\includegraphics[width=0.142\textwidth]{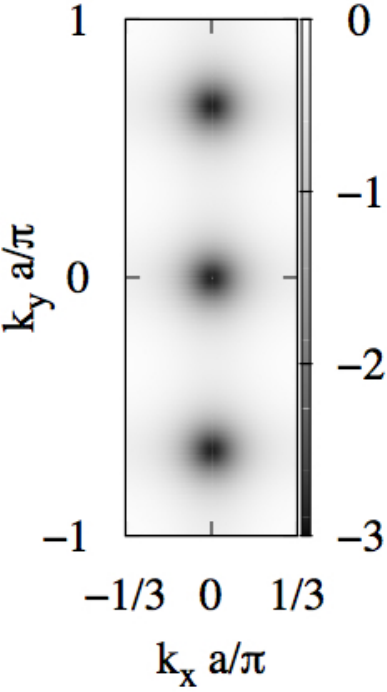}}
\caption{Normalized Berry curvature $\mbox{\boldmath$\Omega$}_{n}/a^2$ of the three magnetic bands in the $\alpha=1/3$ case. The three panels refer to the corresponding panels in Fig.\ref{fig:band13}. We have numerically checked that upon integration over the whole MBZ, the Chern numbers for the three bands are respectively $C_{n}=-1,2,-1$.}
\label{fig:trecurv}
\end{center}
\end{figure}

Physically, the most striking consequence of Eqs. (\ref{eq:semiclassicalberry}) is that the semiclassical dynamics is not completely encoded in the energy spectrum of the Hamiltonian, but keeps track of the relative phase of the eigenstates via their Berry curvature  \eq{eq:curvberry}. While the Berry curvature identically vanishes for systems with simultaneous time-reversal and spatial inversion symmetries, a non-trivial value of $\mathbf{\Omega}_{n,\kk}$ stems in our case from the presence of the external magnetic field breaking time-reversal. 

A plot of the Berry curvature for the three magnetic bands in the $\alpha=1/3$ case is shown in Fig.\ref{fig:trecurv}: its general structure reflects the $q$-fold degeneracy of the energy bands in the $\hat{k}_y$-direction. Thanks to the inversion symmetry of the system, the curvature is symmetric under $\textbf{k}\rightarrow -\kk$. Finally, the sum of the Berry curvature of the different bands at the same point $\kk$ in momentum space identically vanishes throughout the whole MBZ.

The most celebrated consequence of the Berry curvature in electron transport is perhaps the integer quantum Hall effect, where the quantized value of the Hall conductivity  can be related to the {\it Chern number} of the occupied bands, an integer-valued topological invariant proportional to the integral of the Berry curvature over the magnetic Brillouin zone~\cite{thouless}. Other examples of electronic systems displaying bands with non-trivial Berry curvature are ferromagnetic Fe crystals \cite{yao} or graphene single layers with a staggered potential breaking the spatial inversion \cite{zhou,cooper}.

\section{Elementary derivation of the anomalous Berry velocity}

Before proceeding, it is worth presenting an elementary derivation of the semiclassical equations of motions that does not rely on the Lagrangian formalism used in the original works~\cite{niu,sundaram}. For simplicity, we restrict to the simplest case of a weak and constant electric field $\EE$. With a suitable choice of gauge, both the magnetic and the electric fields can be included in a (slowly varying) vector potential 
\begin{equation}
\textbf{A}(\rr,t)=\textbf{A}_{\textbf{B}}(\rr)-\EE \, t,
\label{eq:A_time}
\end{equation}
the first constant term $\textbf{A}_{\textbf{B}}(\rr)$ referring to the constant magnetic field, $\nabla\times \AAA_{{\mathbf B}}={\mathbf B}$. The system Hamiltonian $\bar{H}(t)$ will have again the same form \eq{eq:tbh} but the hopping phases defined by \eq{eq:phase} in terms of the vector potential \eq{eq:A_time} will be now time-dependent. 
The Hamiltonian $\bar{H}(t)$ has the same magnetic translation symmetry properties as the Hamiltonian \eq{eq:tbh} and the eigenstates can be classified in terms of the momentum $\kk$ within the MBZ. Provided the temporal variation $\AAA(t)$ is slow enough, the adiabatic theory guarantees that no inter-band transition will occur. Thanks to the magnetic translational symmetry of $\bar{H}$, each $\kk$ component will be decoupled from all others and will just acquire a $\kk$-dependent phase factor.

Under this approximation, the semiclassical state will be written in the form 
\begin{equation}
 |W(t)\rangle=\int \upd^2\textbf{k} \,w(\textbf{k})\,e^{i\theta(\kk,t)}\,|\bar{\psi}^{(t)}_{n,\textbf{k}}\rangle\,,
 \label{eq:W}
\end{equation}
where $w(\kk)$ is the initial momentum-space wavefunction and $|\bar{\psi}^{(t)}_{n,\textbf{k}}\rangle$ is the eigenstate of the Hamiltonian $\bar{H}(t)$ at a wavevector $\kk$ on the magnetic band $n$ of interest at time $t$. During time-evolution, the eigenfunctions of the Hamiltonian acquires a $\kk$-dependent phase $\theta(\kk,t)$, according to Berry's adiabatic theorem, satisfying
\begin{equation}
\hbar\dot \theta(\kk,t)= -\bar{\mathcal{E}}^{(t)}_{n,\kk} +\hbar  \bar{\mathcal{A}}_{n,\kk}^{(t)}\,,
\label{eq:evol_theta}
\end{equation}
in terms of the eigenenergy $\mathcal{E}^{(t)}_{n,\kk}$ of $\bar{H}(t)$ at the given time $t$ and a Berry term
\begin{equation}
\bar{\mathcal{A}}_{n,\kk}^{(t)}= i \langle \bar{\psi}^{(t)}_{n,\textbf{k}} | \frac{\partial \bar{\psi}^{(t)}_{n,\textbf{k}}}{\partial t} \rangle.
\label{Berry_t}
\end{equation}
By eliminating the time-dependent component of the vector potential with a suitable gauge transformation, one can easily see that the eigenstates of $\bar{H}(t)$ at wavevector $\kk$ coincide with the ones  of the original Hamiltonian \eq{eq:tbh} at wavevector $\kk+e\EE t$, that is $\bar{\mathcal{E}}^{(t)}_{n,\kk} = \mathcal{E}^{(t)}_{n,\kk+e\EE t}$ and
$|\bar{\psi^{(t)}}_{n,\kk}\rangle = |\psi^{(t)}_{n,\kk+e\EE t}\rangle $. As a result, the Berry term \eq{Berry_t} is related to the usual Berry connection on the MBZ by
$ \bar{\mathcal{A}}^{(t)}_{n,\kk} = e\EE \cdot \mathcal{A}_{n,\kk + e\EE t}$.

\begin{figure}[hbtp]
\centering
\includegraphics[width=0.43\textwidth]{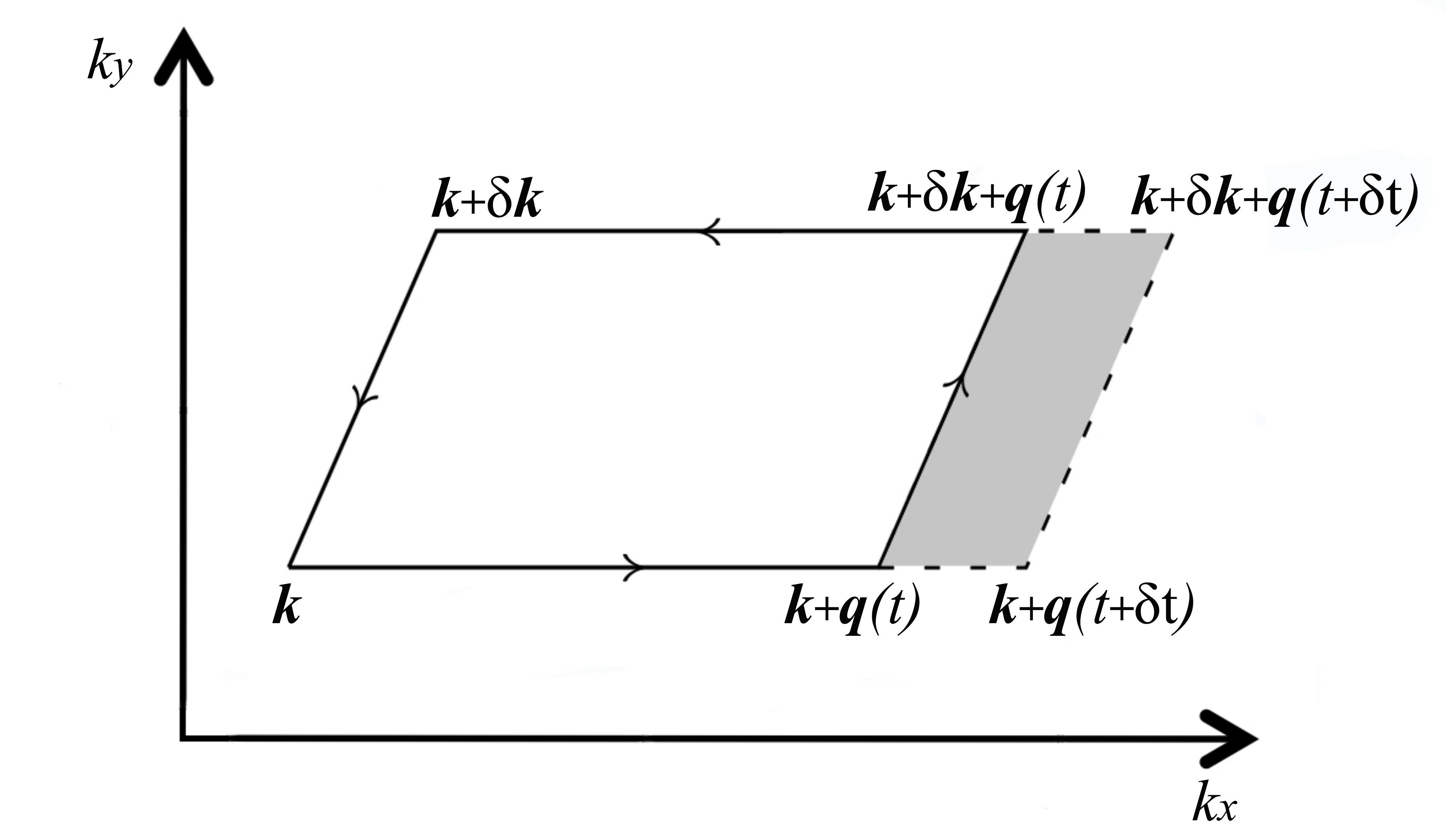}
\caption{Sketch of the integration paths involved in the time evolution of the phase $\theta(\kk,t)$. We indicate $\textbf{q}(t)=e\textbf{E}t.$}
\label{fig:rettangolo}
\end{figure}

To obtain the evolution equation for the real space position $\rr_c$ of the wavepacket, it is then enough to apply the stationary-phase approximation to the wavefunction \eq{eq:W} using the form \eq{eq:evol_theta} of the phase. Under this approximation, the position $\rr_c$ is determined by the $\kk$-gradient of the phase $\theta(\kk,t)$ as
\begin{equation}
 \rr_c = - \nabla_\kk \left[ \textrm{Arg}[w(\kk)]+ \theta(\kk,t) + \mathcal{A}_{n,\kk +e \EE t} \right]\,,
 \label{eq:r_c}
\end{equation}
note in particular the non-trivial term proportional to the Berry connection $\mathcal{A}$ that accounts for the $\kk$-dependence of the phase of the Bloch eigenstate. A temporal derivative of \eq{eq:r_c} gives the equation for the velocity $\upd \rr_c/\upd t$,
\begin{equation}
\hbar \dot \rr_c= \hbar \vv^{\rm gr}_{n,\kk} -e \EE \times \mathbf{\Omega}_{n}(\kk)\,.
\end{equation}
The first term is a direct consequence of the definition $\vv^{\rm gr}_{n,\kk}=\frac{1}{\hbar}\,\nabla_\kk \mathcal{E}_{n,\kk}$. 
The second term can be understood from the sketch in Fig.\ref{fig:rettangolo}: the difference of the phases $\theta$ acquired after a time $t$ by the components of wavevectors $\kk$ and $\kk+\delta \kk$ is given by the difference of the integrals of the Berry connection $\mathcal{A}$ taken along the two horizontal sides of the parallelogram. The variation of the Berry connection term in \eq{eq:r_c} corresponds instead to the difference of the integral of the Berry connection taken along the two shorter sides. The sum of all these contributions recovers the time-derivative of the integral of the Berry connection around the whole parallelogram, that is the (normalized) flux of the Berry curvature through the infinitesimal gray-shaded area, $ \frac{1}{\delta t}\,[ \EE \, \delta t \times \delta \kk]\cdot \mathbf{\Omega}_{n}(\kk)$.

\section{Magnetic Bloch oscillations}

We can now proceed to the main subject of this Letter, that is the effect of the strong magnetic field on the {\em Bloch oscillations} of a quantum particle subject to a uniform and time-independent force.
Starting from the semiclassical Eq.(\ref{eq:semiclassicalberry}a), the oscillation frequency along each principal direction $\hat{x}, \hat{y}$ of the lattice is proportional to the corresponding component ${e}E_{x,y}$ of the applied force,  $\omega_{B}^{(x,y)}= {e} E_{x,y}qa /2\pi \hbar$: the momentum $\textbf{k}$ moves around the MBZ at a constant speed set by the applied force. The factor $q$ in the expression for $\omega_B^{(x,y)}$ is due to the reduced size of the MBZ on the $\hat{x}$-direction, and to the shorter periodicity of the bands along the $\hat{y}$-direction that is visible in Fig.\ref{fig:band13}.

Under the same semiclassical approximation, the spatial motion is ruled by Eq.(\ref{eq:semiclassicalberry}b): in absence of magnetic field, the Berry curvature vanishes $\mathbf{\Omega}(\bf{k})=0$ and the particle trajectory is fully determined by the group velocity. In the simplest case where the force is directed along a principal axis and the momentum starts from a high symmetry point like $\kk=0$, the BO have the usual back-and-forth shape along the direction of the force and their spatial amplitude is $\Delta x = \Delta \mathcal{E} / |e\textbf{E}|$ where $\Delta \mathcal{E}$ is the width of the energy band explored by the particle momentum along its path across the MBZ. In the general case, the complicate shape of the band energy $\mathcal{E}_{n,\mathbf{k}}$ across the MBZ leads to complex trajectories in real space, which do not typically close on themselves and can show lateral drifts~\cite{kolovsky}.

\begin{figure}[hbtp]
\centering
\includegraphics[width=0.45\textwidth]{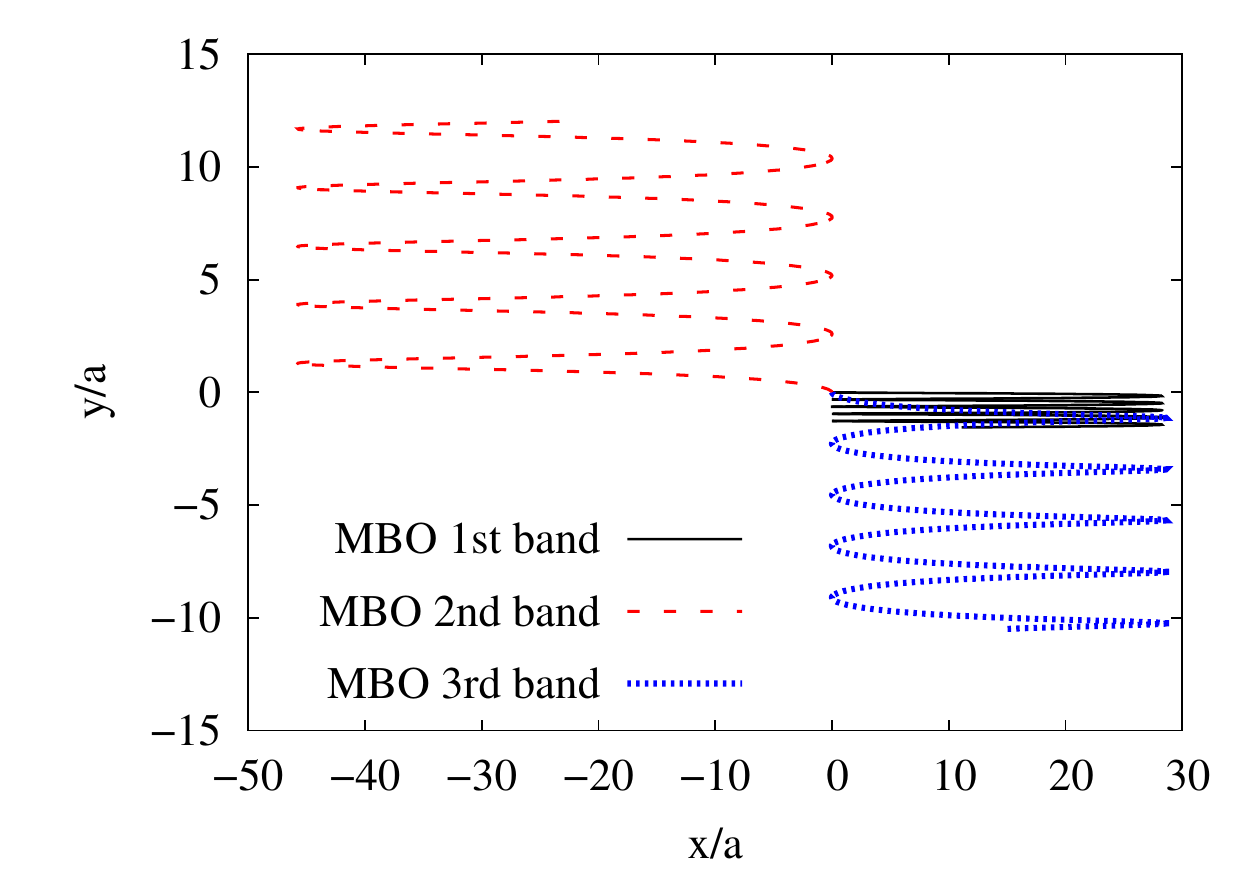}
\caption{Trajectory of the magnetic Bloch oscillations (MBO) in the $\alpha=1/3$ case. The applied electric field is in the $\hat{x}$ direction and has a strength $0.01\,J/a$. The starting point of the MBO is at the origin of both real and reciprocal spaces, $\textbf{r}=0$, $\textbf{k}=0$.}
\label{fig:botutte}
\end{figure}

The situation is far more interesting in the presence of a finite magnetic field $\alpha>0$. To concentrate on the physics of interest, we again start from the high symmetry $\textbf{k}=0$ point in $\textbf{k}$-space and we direct the applied force along the symmetry direction $\hat{x}$: as a result, the wave vector $\textbf{k}$ sweeps the MBZ along the $k_{y}=0$ line and the group velocity remains by symmetry always directed along $\hat{x}$. Still, as it is illustrated in Fig.\ref{fig:botutte} the particle motion from the initial position $\rr=0$ shows a non-vanishing drift along the $\hat{y}$-direction due to the Berry curvature in addition to the periodic Bloch oscillation along $\hat{x}$: the different curves show these magnetic Bloch oscillations (MBO) for each of the three magnetic bands of the $\alpha=1/3$ case, whose Berry curvatures differ in both magnitude and sign. While the average speed of the lateral drift grows with the applied force $e\EE$, the amplitude of the periodic motion is inversely proportional to $e\EE$. 

Note that for arbitrary initial wavevector $\kk$, the trajectory of the MBO can be much more complicate, which makes it difficult to isolate the contribution of the Berry curvature. A clever time-reversal protocol to overcome this difficulty and obtain a complete mapping of the Berry curvature over the whole MBZ was proposed in~\cite{cooper}.

\section{Classical Hall effect in the continuum limit}

It is interesting to see how the classical Hall effect for a charged particle in free space under crossed electric and magnetic fields is recovered in our formalism. To this purpose, one has to let the lattice spacing $a\to 0$ at constant values of the effective mass $m^{*}=\hbar^{2}/2Ja^2$ and of the applied electric $\bf{E}$ and magnetic $\bf{B}$ fields. In this limit, the discreteness of the lattice must disappear and the particle trajectory must recover the textbook cycloidal shape that results from the superposition of a circular motion at the cyclotron frequency $\omega_c={e}B/m$ and a uniform drift at a Hall velocity ${\bf v}_{H}=-{\bf E}\times {\bf B}/B^2$. In a most important limiting case, the circular motion degenerates into a point and one is left with only the uniform drift at the Hall velocity ${\bf v}_H$.

\begin{figure}[hbtp]
\centering
\includegraphics[width=0.46\textwidth]{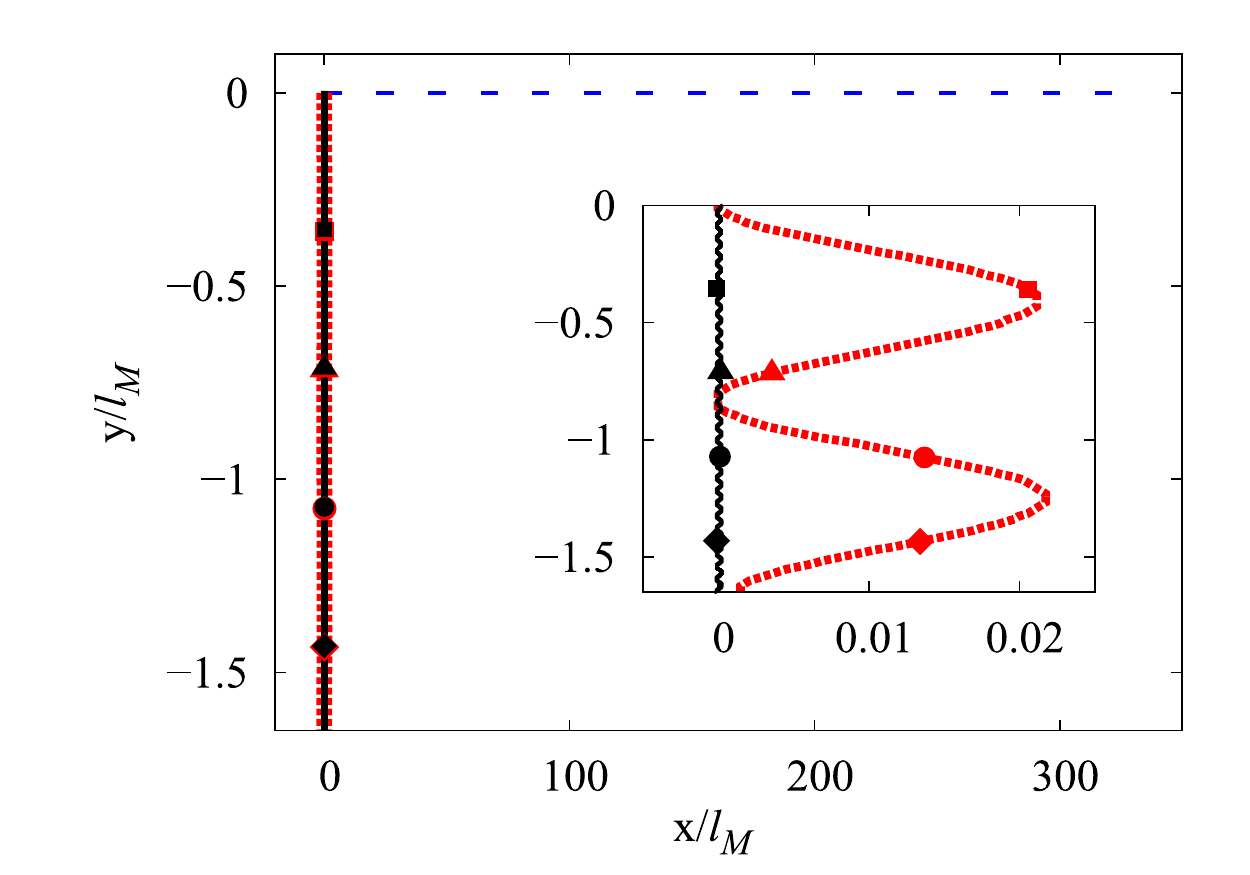}
\caption{Real-space trajectories (measured in units of the magnetic length $l_{M}=\sqrt{\hbar/eB}$) of a particle under crossed electric and magnetic fields for a magnetic field strength $\alpha=1/9$. The black solid line is the classical trajectory in free space for an initial condition such that the circular component of the motion degenerates to a single point. The red dashed line is the MBO in the lowest magnetic band. Corresponding symbols on top of each trajectory indicate the positions at corresponding times. The blue dashed line is the BO trajectory for vanishing magnetic field $\alpha=0$. The inset is a magnified view of the region close to $x=0$.
}
\label{fig:matching}
\end{figure}

In Fig.\ref{fig:matching} we show how this behavior is recovered in our quantum mechanical lattice model.
In the $a\to 0$ limit at constant $\bf{B}$, the magnetic flux per unit cell $\alpha\to 0$, so that the lowest magnetic energy bands flatten and tend to a series of equispaced Landau levels separated by the cyclotron frequency $\omega_c$. In particular, for $\alpha=1/q$ it is known that the Chern number of the lowest band is $C_0=-1$~\cite{goldman}. For large values of $q$ we have observed, in agreement with~\cite{Zak_solid}, that also the Berry curvature of the first band tends to be uniformly distributed across the whole MBZ with $\mathbf{\Omega}_{0}(\kk)\simeq -\hbar \hat{z}/eB$.
Combining all these results, the flatness of the bands reflects into a very small amplitude of the MBO and the Berry curvature term recovers the drift at the Hall velocity ${\bf v}_{H}$: it is quite remarkable to see that to recover the free space classical Hall effect in the quantum mechanical lattice model one has to properly keep into account the Berry curvature!

The small amplitude MBO that are visible in the inset are due to the exponentially small, but still finite width of the magnetic bands: their frequency is determined by the strength $eE$ of the applied force and they are distinct from the cyclotron motion of classical mechanics, whose frequency is instead determined by the magnetic field $B$. For this latter to be recovered in the magnetic band picture, one should in fact take as the initial state of the particle a linear superposition of the different bands, that in the small flux limit are indeed spaced by the cyclotron frequency $\omega_c$.

\section{Conclusions and experimental perspectives}

In this Letter we have given a unified and elementary presentation of the semiclassical theory of transport and of its predictions for the influence of the non-trivial topology of energy bands on the motion of quantum particles in a periodic potential~\cite{niu,sundaram}. The general concepts have been applied in a study of the Bloch oscillations performed by a quantum mechanical particle propagating in a two-dimensional periodic lattice in the presence of a strong (synthetic) gauge field and subject to a uniform force: in addition to the usual periodic motion along the force direction, a lateral drift appears proportional to the Berry curvature. Besides matter wave systems~\cite{dudarev,cooper,modugno_berry}, promising alternative candidates to observe these effects are photonic lattices of coupled waveguides or cavities in which huge synthetic gauge fields for photons have been recently produced~\cite{rechtsman,hafezi} and Bloch oscillations under the effect of a slow spatial gradient of the cavity/waveguide size observed~\cite{pertsch,morandotti,sapienza}. In the cavity case of~\cite{hafezi}, the BO can be followed by monitoring the real-time redistribution of the photonic wavepacket among the cavities after a coherent pulsed excitation. In the waveguide case of~\cite{rechtsman,pertsch,morandotti}, a suitably shaped continuous-wave beam has to be sent onto the waveguide array to selectively excite the initial state; then, the progress of BO is visible from the transverse motion of the beam spot during propagation in space.

\acknowledgments
We thank \textsc{R. O. Umucalilar} for useful discussions. This work has been supported by ERC through the QGBE grant and by Provincia Autonoma di Trento.

\bibliographystyle{eplbib}
\bibliography{biblio}

\end{document}